\documentclass[journal]{IEEEtran}
\usepackage{amsmath}
\usepackage{amssymb}
\usepackage{lipsum}
\usepackage{graphicx}
\usepackage[justification=centering]{caption}
\usepackage{epstopdf}
\usepackage{tabularx}
\usepackage{algorithm}
\usepackage{algorithmic}
\usepackage{multirow}
\usepackage{booktabs}

\usepackage{hyperref}

\ifCLASSINFOpdf
\else
\fi
\begin{document}
%
\title{A Survey on Personality-Aware Recommendation Systems}
%
%

\author{Sahraoui Dhelim, Nyothiri Aung, Mohammed Amine Bouras, Huansheng Ning and Erik Cambria.
\thanks{Sahraoui Dhelim \and Nyothiri Aung \and Mohammed Amine Bouras \and Huansheng Ning are with School of Computer and Communication Engineering, University of Science and Technology Beijing, 100083, Beijing, China.}
\thanks{Erik Cambria is with Nanyang Technological University, Singapore}
\thanks{Corresponding author: Huansheng Ning (ninghuansheng@ustb.edu.cn).}
}
%
%

\markboth{Submitted to Artificial Intelligence Review.}%
{Dhelim \MakeLowercase{\textit{et al.}}: Bare Demo of IEEEtran.cls for IEEE Journals}
%



\maketitle

\begin{abstract}

With the emergence of personality computing as a new research field related to artificial intelligence and personality psychology, we have witnessed an unprecedented proliferation of personality-aware recommendation systems. Unlike conventional recommendation systems, these new systems solve traditional problems such as the cold start and data sparsity problems. This survey aims to study and systematically classify personality-aware recommendation systems. To the best of our knowledge, this survey is the first that focuses on personality-aware recommendation systems. We explore the different design choices of personality-aware recommendation systems, by comparing their personality modeling methods, as well as their recommendation techniques. Furthermore, we present the commonly used datasets and point out some of the challenges of personality-aware recommendation systems.

\end{abstract}

\begin{IEEEkeywords}

Recommendation system, Personality computing, personality traits, Big-five model, personality-aware, social computing, collaborative filtering.

\end{IEEEkeywords}

%
\IEEEpeerreviewmaketitle

\section{Introduction}
%
%
%
%
\IEEEPARstart{P}{ersonality} Computing is the interdisciplinary study field that focuses on the integration of personality psychology theories with computing systems. It has been proven that leveraging personality theories could help to tackle some of the well-known problems in computer science. Personality computing has been applied in many domains and research directions, and the number of scientific publications within the scope of personality computing has dramatically increased within the last decade. The integration of user personality traits into the computing system has created new research directions, such as automatic personality recognition, and helped to accelerate existing research directions as well, such as recommendation systems, and human-robot interaction research. Personality computing has enabled recommendation systems to understand the users' preferences from a different perspective. A new type of recommendation system that leverages the user's personality trait to improve the recommendations had emerged. This group of systems is known as Personality-aware recommendation systems. This new type of recommendation systems has proven effective in solving the problem of conventional recommendation systems. Such as the cold-start problem, when the system does not have much data about the preferences of the user, free-riders problem and the data sparsity problem, to name a few.

In the recent few years, we have witnessed a rapid proliferation of personality-aware recommendation systems. While all of these recommendation systems incorporate the user's personality traits in the recommendation process, however, these systems use different recommendation techniques, and they are designed for different recommended content. Therefore, in this paper, we conduct a comprehensive review of the literature of personality-aware recommendation systems.  Few works surveyed some research direction in the field of personality computing. In 2014, Vinciarelli and Mohammadi \cite{Vinciarelli2014} surveyed the publications that used the user's personality in computing systems, and they coined the term Personality Computing. In 2017, Kaushal and Patwardhan \cite{Kaushal2018} surveyed the literature on automatic personality recognition from online social networks. Similarly, in 2019, Mehta \textit{et al.}. \cite{Mehta2020} surveyed the literature on deep-learning-based personality automatic personality recognition. However, as far as we know, we are the first who survey the literature of personality-aware recommendation systems. In Tables \ref{surveys_table}, we list some of the recent surveys in the field of personality computing, along with their focus scope and publication year.

The remainder of this paper is organized as follows:

In Section 2, we show the main differences between conventional recommendation systems and personality-aware recommendation systems by explaining the main component of the latter. While in Section 3, we systematically classify the existing personality-aware recommendation system based on the used recommendation technique. In Section 4, we review some of the works that proposed personality-aware recommendation systems in the last few years. Whereas in Section 5, we present some of the commonly used datasets and benchmarks related to personality-aware recommendation systems. In Section 6, we discuss some of the challenges that face personality-aware recommendation systems and also list some of the open issues and research challenges. Finally, we conclude this survey in Section 7.

\begin{table*}[]
	\centering
	\caption{Related surveys and reviews}
	\label{surveys_table}

\begin{tabular}{|m{0.7in}|m{1in}|m{2in}|m{0.3in}|m{2in}|} \hline 
	\textbf{Research field} & \textbf{Publication} & \textbf{Scope description} & \textbf{Year} & \textbf{Note} \\ \hline 
	\multirow{1}{*}{\parbox{1\linewidth}{\vspace{1cm}Personality computing}}  
	& Vinciarelli and Mohammadi \cite{Vinciarelli2014} & A general survey on personality computing & 2014 & Vinciarelli and Mohammadi coined the term ``Personality Computing'' \\\cline{2-5}
	& Wright \cite{G.C.Wright2014} & Commentary about \cite{Vinciarelli2014} & 2014 & Wright explain his perspective about personality computing as a personality psychologist \\\cline{2-5}
	& Vinciarelli and Mohammadi \cite{Vinciarelli20142} & A complement of \cite{Vinciarelli2014} by the same authors. & 2014 & Vinciarelli and Mohammadi replied on the commentary of Wright \cite{G.C.Wright2014}, and discussed the future direction of Personality Computing. \\\cline{1-5}
\multirow{1}{*}{\parbox{1\linewidth}{\vspace{2.5cm} Automatic \\personality\\ recognition}} 
 & Jacques \textit{et al.} \cite{SilveiraJacquesJunior2019} & A survey on vision-based personality detection & 2019 & The authors have surveyed only image-based and video-based personality detection \\\cline{2-5} 
	& Kaushal \textit{et al.} \cite{Kaushal2018} & A survey on user personality detection from online social networks & 2018 &  \\\cline{2-4} 
	& Mehta \textit{et al.} \cite{Mehta2020} & A survey on deep learning based personality detection. & 2019 &  \\\cline{2-4} 
	& Bhavya \textit{et al.} \cite{Bhavya2020} & A review on deep learning based personality detection from online social networks. & 2019 & These works surveyed textual, as well as non-textual (image, video, voice) personality detection. \\ \cline{2-4} 
	& Kedar \textit{et al.} \cite{Kedar2015} & A review on various approaches used for personality assessment & 2015 &  \\ \cline{2-4} 
	& Finnerty \textit{et al.} \cite{Finnerty2016} & A review on the data sources and the features used for automatically infer personality & 2016 &  \\ \cline{2-4}
	& Azucar \textit{et al.} \cite{Azucar2018} & Review on Big 5 personality traits from digital footprints on social media & 2017 &  \\ \cline{2-5} 
	& Dandannavara \textit{et al.} \cite{Dandannavar2018} & A review on text-based personality prediction from online social networks & 2019 & The authors have reviewed only text-based personality detection \\ \hline 
	\multirow{1}{*}{\parbox{1\linewidth}{\vspace{0.5cm} Personality in human-robot interaction}}
	 & Santamaria \textit{et al.} \cite{Santamaria2017} & A review on research methods for measuring personality in human-robot interaction & 2017 &  \\ \cline{2-4}
	& Robert \cite{robert2018personality} & A review on personality in human-robot interaction & 2019 &  These works focus on robots' artificial personality traits and the application of personality computing in human-robot interaction  \\ \cline{2-4} 
	& Robert \textit{et al.} \cite{Robert2020} & A review on personality in human-robot interaction & 2020 & \\ \cline{2-4} 
	& Mou \textit{et al.} \cite{Mou2020} & A review on the personality of robots & 2020 &  \\ \hline 
\end{tabular}

\end{table*}

\section{Personality-aware recommendation systems}
\begin{figure}[!htbp]
	\centering
	\includegraphics[width=\columnwidth]{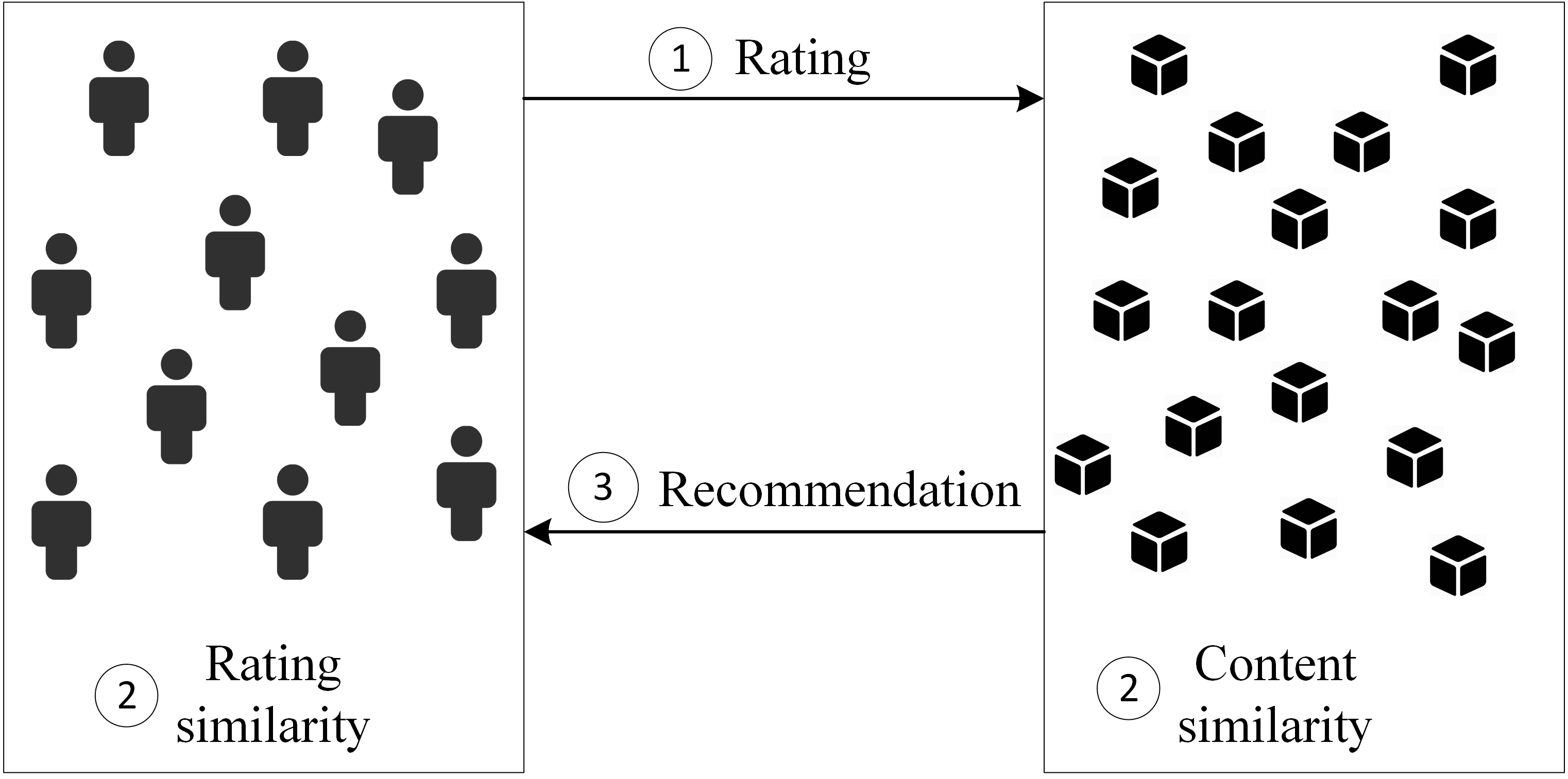}
	\caption{Conventional recommendation systems}
	\label{conv_rs}
\end{figure}

\begin{figure}[!htbp]
	\centering
	\includegraphics[width=\columnwidth]{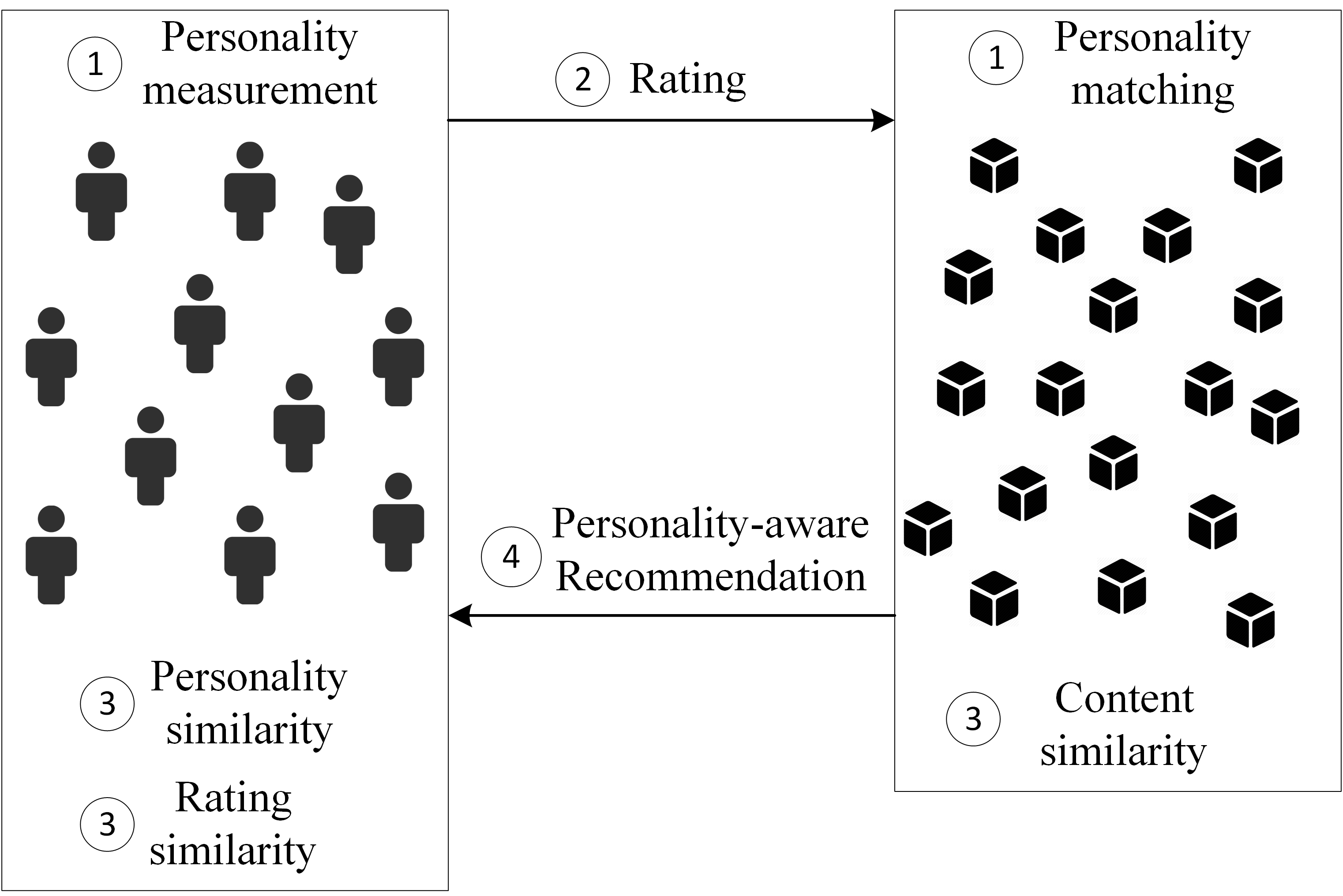}
	\caption{Personality-Aware recommendation systems}
	\label{pers_rs}
\end{figure}

Historically, recommendation systems are divided into three main categories, collaborative filtering approaches, content filtering approaches and hybrid filtering approaches. Collaborative filtering is inspired by the fact that ``people who agree on the past, probably will agree in the future''. In practice, in order to recommend new items to a given user $u_x$, collaborative filtering systems determine a group of users that have a similar rating with user $u_x$, these users are called the neighbors of user $u_x$. After finding the group of neighbors, the system finds the set of items that share a high rating among these neighbors, and subsequently recommend these items to user $u_x$. While content filtering approaches, compute the similarity between previous matched items and the suggested items, regardless of the neighbors' ratings. Finally, hybrid approaches use a combination of these two techniques. Similar to the conventional recommendation systems, personality-aware recommendation systems also use similar recommendation techniques, the only difference is that they add the user's personality information in the recommendation process. In Figure \ref{conv_rs} and Figure \ref{pers_rs}, we show the main differences between conventional and personality-aware recommendation systems. Conventional recommendation systems mainly have three stages. Firstly, the rating phase, where the user expresses her interests by rating some items. The second stage is the filtering phase, either collaborative filtering, content filter or hybrid filtering as mentioned above. Finally, at the recommendation phase, the system recommends the items yielded by the filtering phase. 

As shown in Figure \ref{pers_rs}, personality-aware recommendation systems add two more phases before the rating phase and change the filtering stage as well. Personality measurement phase, the system assesses the personality type of the users using a personality assessment questionnaire that the users need to answer during registration, or by applying an automatic personality recognition scheme on the users' previously available data, such as online social network data. While in the personality matching phase, the system tries to match the user personality type with relevant items, this could be done either using lexical matching by linking the textual description of the items with the associated personality types, or using a fine-grained rules that can match items with personality types. It is worth noting that at personality matching phase, the system does not have any information about the users' ratings, which help to alleviate the effects of cold-start problem, one of the most challenging problems in the literature of recommendation systems. Personality-aware recommendation systems also change the filtering phase, by incorporating the personality information in the similarity measurement to determine the neighbors of each user.

\subsection{Personality measurement}
The personality measurement is the most important phase of personality-aware recommendation system, as any misidentification of the user personality type could negatively influence the accuracy of the recommendation system. In personality computing, there are two main methods for personality measurement, personality assessment questionnaires and automatic personality recognition (APR). Generally, questionnaires based personality measurement is more accurate than APR. However, in the context of personality computing, APR are relatively easier to conduct, as they can be applied to the user's existing data, without the need to burden the user with long questionnaires. In this subsection, we discuss the different personality assessment questionnaires and APR techniques and classify some of the existing personality-aware recommendation systems based on the their personality assessment method.

\subsubsection{Personality assessment questionnaires}
In the study field of personality psychology, self-report personality questionnaires have been widely used to reveal personality differences among individuals. Questionnaires in which people estimate their character and behaviors are the most commonly used mean for personality assessment. The answers are typically in the format of five-level Likert scale (strongly agree, agree, neither agree or disagree, disagree and strongly disagree). 

There are many personality questionnaires of various lengths (number of items). The widely used long questionnaires include NEO Five-Factor Inventory (NEO-FFI, 60 items) \cite{Aluja2005}, NEO-Personality-Inventory Revised (NEO-PI-R, 240 items) \cite{costa2008revised}, International Personality Item Pool (IPIP-NEO, 300 items) and the Big-Five Inventory (BFI, 44 items) \cite{Rammstedt2007}. In the context of personality-aware recommendation system, these questionnaires could be directly solicited from the users during registration. Filling a long questionnaire is a time-consuming task, the user might get bored and may not fill the questionnaire carefully, which could lead to an irrelevant recommendation in the future. Therefore, short questionnaires \cite{Rammstedt2007,Gosling2003,topolewska2014short} (5-10 items) are preferred in personality-aware recommendation systems, as these questionnaires are much easier to fill. The most prominent short questionnaires are BFI-10 (a short version of BFI with 10 items) and Ten-Item Personality Inventory (TIPI, 10 items) \cite{Gosling2003}. Table \ref{bfi_table} shows the items of BFI-10.

\begin{table}[]
	\centering
	\caption{The BFI-10 Personaltiy questionnaire}
	\label{bfi_table}
\begin{tabular}{|p{0.2in}|p{1.5in}|p{0.8in}|} \hline 
	\textbf{Item} & \textbf{Question} & \textbf{Dimension} \\ \hline 
	1 & I am outgoing, sociable & Extraversion \\ \hline 
	2 & I get nervous easily & Neuroticism \\ \hline 
	3 & I tend to be lazy & Conscientiousness \\ \hline 
	4 & I have an active imagination & Openness \\ \hline 
	5 & I am reserved & Extraversion \\ \hline 
	6 & I am generally trusting & Agreeableness \\ \hline 
	7 & I have few artistic interests & Openness \\ \hline 
	8 & I tend to find fault with others & Agreeableness \\ \hline 
	9 & I do a thorough job & Conscientiousness \\ \hline 
	10 & I am relaxed, handle stress well & Neuroticism \\ \hline 
\end{tabular}
\end{table}

There are two main drawbacks of self-assessment questionnaires. The first limitation is the self-bias problem \cite{PEDREGON2012213}, when the subjects tend to give the wrong answer to some of the undesired social characteristics in certain circumstances. For example, when answering a recruitment personality questionnaire, most of the subjects gives inaccurate answers to questions like ``I get nervous easily'', ``I tend to be lazy'', because these are undesired characters in employees. The self-bias does not affect personality-aware recommendation systems, because users have no benefit in misleading the system. The second drawback is known as the reference-group effect \cite{youyou2017birds}, in which the answers given by the subject is relative to his surrounding environment. For example, an introvert engineer might think he is extravert if he is surrounded by a group of even more introvert engineer friends. In Table \ref{questionnaires} we list the personality assessment questionnaires that were used in the recent personality-aware recommendation system.

\begin{table}[]
	\centering
	\caption{Questionnaire types in personality-aware recommendation systems}
	\label{questionnaires}
\begin{tabular}{|m{0.9in}|m{1in}|m{0.4in}|} \hline 
	\textbf{Recommendation system} & \textbf{Personality assessment questionnaire} & \textbf{Item count} \\ \hline 
	\cite{Karumur2018,Nguyen2018,Zheng2020,Buettner2017,Wu2019a,berkovsky2017recommend,Hu2010a,Elahi2013,hu2010using,Hu2011,Dhelim2020KBS,sun2018,Liu2020} & TIPI  & 10 \\ \hline 
	\cite{Marko2010,Felfernig2012,cantador2014exploitation,Neehal2019,5050recommend}  & IPIP 50-version & 50 \\ \hline 
	\cite{Wu2018,Balakrishnan2018,Schedl2016,Ting2018,sofia2016investigating,ferwerda2016influence} & BFI & 44 \\ \hline 
	\cite{YusefiHafshejani2018,Shayegan2020} & NEO-FFI & 60 \\ \hline 
	\cite{Subramanian2018,Lepri2012,zen2010space,Subramanian2013} & Big-five marker scale (BFMS) \cite{perugini2002analyzing} & 100 \\ \hline 
	\cite{roffo2016towards,guntuku2015,enlighten149660} & BFI-10 & 10 \\ \hline 
	\cite{ning2019personet} & IPIP-NEO-60  & 60 \\ \hline 
	\cite{Mukta2016} & IPIP 44-version & 44 \\ \hline 
	\cite{Fernandez_Tobias2016}  & IPIP 336-version & 336 \\ \hline 
	\cite{Youyou2017} & IPIP 100-version & 100 \\ \hline 
	\cite{Braunhofer2015} & FIPI & 5 \\ \hline 
\end{tabular}
\end{table}

\subsubsection{Automatic personality recognition}

The assessment of users' personality using a questionnaire is not possible in certain circumstances, for example when analyzing an existing anonymous dataset, or when filling a personality questionnaire is not convenient. APR could be used to solve this dilemma. APR is the process of mapping the data related to a subject to a personality score that represents the personality type of that subject. In the context of user personality from online social network data, APR schemes are generally divided into three classes. Text-based APR, where the source data is in text format such as social media posts or tweets. Multimedia-based APR, where the source data is an image, voice or video, such as social media profile photos. And finally, behavior-based APR, where the source data represent a set of behavioral patterns of the user, such as gaming behaviors or browsing behaviors. Text-based APR generally has higher accuracy than multimedia-based APR and behavior-based APR.

Text-based APR is inspired by the fact that some language psychology theories claim that the choice of words can reveal some psychological states such as emotions and personality traits of the subject \cite{Hirsh2009}. Therefore, text-based APR analysis the word choice frequency infer the user's personality traits from his social network posts or messages. One of the most common prominent techniques for text-based APR is Linguistic Inquiry and Word Count (LIWC) \cite{Tausczik2010,Mehta2020ICDM}. LIWC categorizes the analyzed text into various psychologically relevant sets known as ``buckets'' like `function words' (e.g., conjunctions, articles, pronouns), `social processes' (e.g., mate, talk, friend) and `affective processes' (e.g., happy, nervous, cried). Following that, LIWC measures the frequency of words in each of these buckets and predicts the personality traits of the subject accordingly. Another famous linguistic database is the Medical Research Council (MRC) psycholinguistics database. Linguistic analysis model like LIWC and MRC have been proven to achieve acceptable accuracy to detect the user's personality traits from its text. For instance, Han \textit{et al.} \cite{Han2020} introduced an APR model based on personality lexicon by analyzing the correlations between personality traits and semantic categories of words, and extract the semantic features of users' microblogs to construct a prediction model using word classification algorithm. On the other hand, multimedia-based APR detects the user's personality traits by analyzing it is related to photos or video and try to associate the features of these data with the facets of personality traits. For instance, users who frequently post photos related to art might achieve a high score openness trait. Li \textit{et al.} \cite{Li2020} introduced a framework that predicts the aesthetics distribution of an image and the Big-Five personality traits of people who like the image. Finally, behavior-based APR detects the user's personality trait by analyzing behavioral patterns and associate them with relevant dominant traits. Annalyn \textit{et al.} \cite{Annalyn2018} studied the relationship content labels ``tags'' generated by users from Goodreads.com, and match it with personality scores collected from Facebook users. Vinciarelli and Mohammadi \cite{Vinciarelli2014} surveyed the literature of APR and classify the reviewed works, and Kaushal \textit{et al.} \cite{Kaushal2018} surveyed APR methods that leverage online social networks as a data source. While Jacques \textit{et al.} \cite{SilveiraJacquesJunior2019} surveyed vision-based APR methods, and recently, Mehta \textit{et al.} \cite{Mehta2020} and Bhavya \textit{et al.} \cite{Bhavya2020} surveyed deep-learning-based APR.  In Table \ref{apr_review}, we summarize some of the key recent APR works that were not covered in these surveys. 

\subsection{Personality matching}
In personality matching phase, the recommendation system computes the matching likelihood between a given user and some items, the matching is computed based only on the personality information of the user and some personality features of the item, such as a product brand in product recommendation or the personality type of actors in the case of movie recommendation. It is worth noting that at this stage the recommendation system does not know the rating information of the user yet, which helps the system to cope with the user cold start problem.

\subsection{Personality filtering}
The primary objective of the filtering phase in the conventional collaborative filtering is to determine the set of neighbors that have similar ratings with the current user, a process known as neighborhood formation. In personality-aware recommendation system, the similarity between the users is computing based on their personality trait similarity or using a hybrid personality-rating similarity measurement, and the resulting set of neighbors are similar in terms of personality traits to the studied user.

\begin{table*}[]
	\centering
	\caption{APR literature review}
	\label{apr_review}
\begin{tabular}{|p{0.6in}|p{0.7in}|p{0.3in}|p{4in}|} \hline 
	\textbf{APR type} & \textbf{Publication} & \textbf{Year} & \textbf{Description} \\ \hline
	\multirow{1}{*}{\parbox{1\linewidth}{\vspace{1.5cm}Text-based APR}}  & Silva \textit{et al.} \cite{Silva2018} & 2018 & Proposed a supervised models for APR of text in Brazilian Portuguese from Facebook posts \\ \cline{2-4} 
	& Han \textit{et al.} \cite{Han2020} & 2020 & Used word embedding techniques and prior-knowledge lexicons to automatically construct a Chinese semantic lexicon suitable for personality analysis. \\ \cline{2-4}  
	& Sun \textit{et al.} \cite{Sun2020} & 2020 & Proposed a model of group-level personality detection by learning the influence from text generated networks \\ \cline{2-4}  
	& Darliansyah \textit{et al.} \cite{darliansyah2019sentipede} & 2019 & Take advantage of Neural Network Language Model for personality detection from short texts by using a unified model that combines a Recurrent Neural Network named Long Short-Term Memory with a Convolutional Neural Network. \\ \cline{2-4}  
	& Santos \textit{et al.} \cite{DosSantos2020} & 2020 & Discussed the effectiveness of using psycholinguistic knowledge in APR, and performed series of individual experiments of APR from Facebook text \\ \hline 
	\multirow{1}{*}{\parbox{1\linewidth}{\vspace{3cm}Multimedia-based APR}} & Li \textit{et al.} \cite{Li2020} & 2020 & Introduced a framework that predicts the aesthetics distribution of an image and Big-Five (BF) personality traits of people who like the image. \\ \cline{2-4}  
	& Kim \textit{et al.} \cite{Kim2018} & 2018 & Used computer vision techniques to detect users personality from their shared pictures. An online survey of 179 university students was conducted to measure user characteristics, and 25,394 photos in total were downloaded and analyzed from the respondents' Instagram accounts. \\ \cline{2-4}
	& Segalin \textit{et al.} \cite{Segalin2017} & 2016 & Used Computational Aesthetics to infer the personality traits of Flickr users from their galleries, their method maps low-level features extracted from the pictures into numerical scores corresponding to the Big-Five Traits, both self-assessed and attributed. \\ \cline{2-4} 
	& Ferwerda \textit{et al.} \cite{Ferwerda2015a} & 2015 & Conducted an online survey, by analyzing 113 participants and 22,398 extracted Instagram pictures, they conclude that there is correlation between distinct picture features and personality traits. \\ \cline{2-4} 
	& Zhu \textit{et al.} \cite{Zhu2020} & 2020 & Introduced an end-to-end weakly-supervised dual convolutional network for personality detection, composed of a classification network and a regression network. The classification network detects personality class-specific attentive image regions. While the regression network is used for detecting personality traits. \\ \cline{2-4} 
	& Guntuku \textit{et al.} \cite{Guntuku2018} & 2016 & Propose a personality traits detection model, and analyzed collection of images and users who tag as favorite on Flickr. \\ \cline{2-4} 
	& Zhang \textit{et al.} \cite{Zhang2020} & 2020 & Proposed PersEmoN, an end-to-end trainable and deep Siamese-like network. PersEmoN is composed of two convolutional network branches, the first for emotion and the second for personality traits. Both networks share their bottom feature extraction module and are optimized within a multi-task learning framework. \\ \hline 
	\multirow{1}{*}{\parbox{1\linewidth}{\vspace{1cm}Behavior-based APR}} & Tadesse \textit{et al.} \cite{Tadesse2018} & 2018 & Analyzed and compared four machine learning models to investigate the relationship between user behavior on Facebook and big-five personality traits. \\ \cline{2-4}
	& Annalyn \textit{et al.} \cite{Annalyn2018} & 2018 & Investigated the relationship between user book preferences by analyzing labels ``tags'' generated by users from Goodreads.com, and match it with personality scores collected from Facebook users   \\ \cline{2-4} 
	& Fong \textit{et al.} \cite{Fong2015} & 2015 & Discussed personality inferences and intentions to befriend based solely on online avatars. \\ \cline{2-4} 
	& Nave \textit{et al.} \cite{Nave2018} & 2018 & Investigated the possibility of personality prediction using musical preferences. Their finding using data of active listening and Facebook likes show that reactions to unfamiliar musical excerpts predicted individual differences in personality. \\ \hline 
\end{tabular}
\end{table*}

\section{ Personality-aware schemes classifications}
\begin{figure}[!htbp]
	\centering
	\includegraphics[width=\columnwidth]{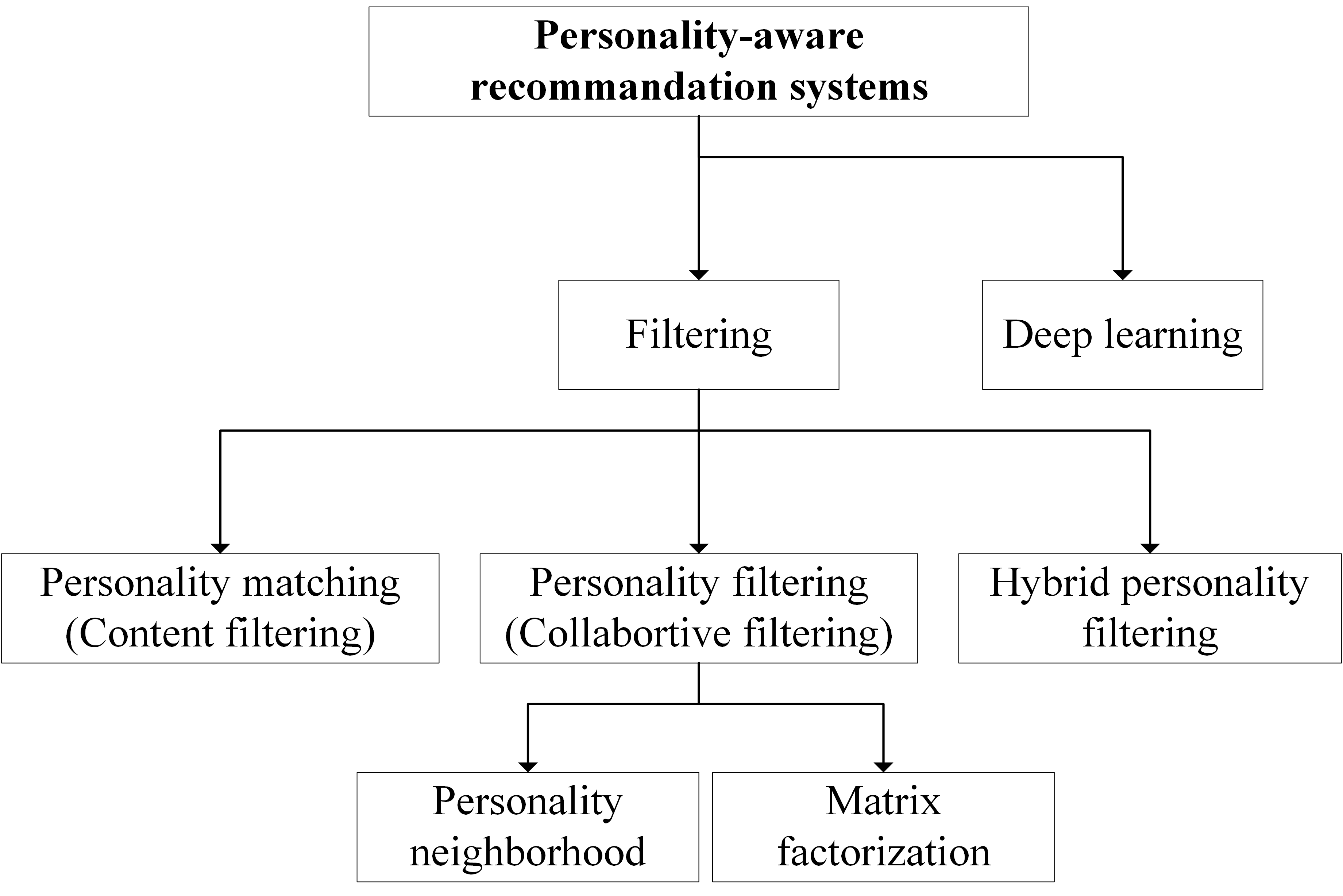}
	\caption{Personality-aware recommendation systems classification}
	\label{classification}
\end{figure}

After the emergence of personality computing, in the last decade, we have witnessed an unprecedented proliferation of personality-aware recommendation systems. These systems use different recommendation techniques, and in some cases, the recommendation process depends on the nature of the recommended content. In this section, we classify the recent personality-aware recommendation system based on the recommendation technique. Personality-aware recommendation systems are roughly divided into four main classes, filtering-based methods and deep-learning-based methods. Figure \ref{classification} shows the classification that we will be using to classify the recent proposed personality-aware recommendation systems. Filtering methods are divided into three classes, personality filtering, personality matching and hybrid filtering.

\subsection{Personality filtering}
Personality-aware recommendation systems that leverage the conventional collaborative filtering technique to filter users with similar personalities are known as personality filtering methods. Personality filtering methods in turn could be further divided into personality-neighborhood methods and matrix factorization methods.
\subsubsection{Personality neighborhood methods}
Personality neighborhood filtering is the most common personality-aware recommendation technique. Typically, the system uses a proximity function that measures the personality similarity to find the personality neighborhood users, and use it to predict future rating accordingly. While there are many proximity functions, the Pearson correlation coefficient is the most commonly used proximity function. Given two users $u_x$ and $u_y$, the rating similarity between them is computed using the function $SimR\left(u_x,\ u_y\right)$ as shown in eq(\ref{equ_SimR}), where $R_x$ and $R_y$ is the sets of the previous rating of user $u_x$ and $u_y$ respectively, and $r_{x,i}$ is the rating of user $u_x$ on item $i$, and $\overline{r_x}$ is the mean rating of user $u_x$.

\begin{equation}
\label{equ_SimR}
\resizebox{0.91\hsize}{!}{$
SimR\left(u_x,\ u_y\right)=\frac{\sum_{i\in R_x\cap R_y}{\left(r_{x,i}-\overline{r_x}\right)\left(r_{y,i}-\overline{r_y}\right)}}{\sqrt{\sum_{i\in R_x\cap R_y}{{\left(r_{x,i}-\overline{r_x}\right)}^2\sum_{i\in R_x\cap R_y}{{\left(r_{y,i}-\overline{r_y}\right)}^2}}}}\  
$}
\end{equation}

Many personality-aware recommendation systems extend this approach to measure the personality similarity between users, as shown in eq(\ref{equ_SimP}), where $\overline{p_x}$ and $\overline{p_y}$ is the average value of the personality traits vector for user $u_x$ and $u_y$ respectively, and $p^i_x$ is the $i^{th}$ trait in the personality traits vector

\begin{equation}
\label{equ_SimP}
\resizebox{0.91\hsize}{!}{$
	SimP\left(u_x,\ u_y\right)=\frac{\sum_i{\left(p^i_x-\overline{p_x}\right)\left(p^i_y-\overline{p_y}\right)}}{\sqrt{\sum_i{{\left(p^i_x-\overline{p_x}\right)}^2\sum_i{{\left(p^i_y-\overline{p_y}\right)}^2}}}}\  
	$}
\end{equation}

However, some other works opted to use other proximity functions to measure the personality similarity between users. In Table \ref{similarity_measures}, we summarize some commonly used personality similarity proximity functions.

After computing the similarity among users and eventually establishing the neighborhood of each user, the prediction score is computed by aggregating the rating of neighborhood users and the similarity with these users. Formally, let $score(u,i)$ denote the predication score that user $u$ will give to item $i$, the prediction score is computed using eq(\ref{equ_score})

\begin{equation}
\label{equ_score}
score\left(u,i\right)=\overline{r_u}+k\sum_{v\in {\mathit{\Omega}}_u}{sim\left(u,v\right)\ (r_{v,i}-\overline{r_v})}
\end{equation}

where $\overline{r_u}$ and $\overline{r_v}$ are the average rating of user $u$ and user $v$ respectively, and $r_{v,i}$ is the rating given by user $v$ to item $i$, and ${\mathit{\Omega}}_u$ are the neighbors of user $u$ that have previously rated item $i$. Different works used a different design of the proximity function that measures the total similarity $sim\left(u,v\right)$. In this regard, there are three main designs. Some works (\cite{Wu2015}) simply use the personality similarity function $simp\left(u,v\right)$ instead of $sim\left(u,v\right)$. While other works (\cite{ning2019personet}) opted to use a combination of the personality similarity function $simp\left(u,v\right)$ and the rating similarity function $simr\left(u,v\right)$. Finally, some other works (\cite{Dhelim2020TCSS,Dhelim2020KBS}) use other social factor similarity functions such as user interests similarity along with the rating similarity and personality similarity. 

\begin{table*}[]
	\centering
	\caption{Personality-based similarity measurement}
	\label{similarity_measures}
\begin{tabular}{|p{0.6in}|m{1in}|m{2in}|p{2in}|} \hline 
	\textbf{Publication} & \textbf{Proximity function} & \textbf{Formula} & \textbf{Note} \\ \hline 
	\cite{Asabere2020,ning2019personet,Xia2017,Asabere2018,YusefiHafshejani2018,Hu2011,sun2018,Recio_Garcia2009,hu2010using} & Pearson correlation coefficient & \parbox{1\linewidth}{\vspace{0.5cm}\resizebox{0.91\hsize}{!}{$SimP \left(u,\ v\right)=\frac{\sum_i{\left(p^i_u-\overline{p_u}\right)\left(p^i_v-\overline{p_v}\right)}}{\sqrt{\sum_i{{\left(p^i_u-\overline{p_u}\right)}^2\sum_i{{\left(p^i_v-\overline{p_v}\right)}^2}}}}$}}  & Used by most of the state-of-the-art personality-aware recommendation systems  \\ \hline 
	\cite{Wu2018,Marko2010} & Normalized Euclidean distance  & \resizebox{0.91\hsize}{!}{$SimP\left(u,\ v\right)=\frac{1}{1+\left(\sqrt{\sum_k{w^2_k{\left(P^k_u-P^k_v\right)}^2}}\right)}$} & $w_k$ is the weight of the k${}^{th}$ personality trait. \\ \hline 
	\cite{Buettner2017} & Euclidean distance  & \parbox{1\linewidth}{\vspace{0.5cm}\resizebox{1\hsize}{!}{$Eucl=\sqrt{{\left(E^{BFI-S}-E^{PPS}\right)}^2+{\left(A^{BFI-S}-A^{PPS}\right)}^2+{\left(C^{BFI-S}-C^{PPS}\right)}^2}$}} & $E^{BFI-S}$ is the extraversion trait value for the user, and $E^{PPS}$ is the extraversion trait value assigned to a product. Similarly A is the agreeableness trait, and C is conscientiousness \\ \hline 
	\cite{Yang2019,Wu2015,Zheng2020,onori2016comparative} & Cosine similarity measure & $SimP\left(u,\ v\right)=\frac{\sum_k{P_{u,k}.P_{v,k}}}{\sqrt{\sum_k{{P_{u,k}}^2}}\sqrt{\sum_k{{P_{v,k}}^2}}}$ & Cosine similarity is a dot product of unit vectors, unlike Pearson correlation, which is cosine similarity between centered vectors.  \\ \hline 
	\cite{Fernandez_Tobias2014} & Spearman's correlation coefficient & \resizebox{0.91\hsize}{!}{$SimP\left(u,\ v\right)=\frac{\sum_k{\left(s_{u,k}-\overline{s_u}\right)\left(s_{v,k}-\overline{s_v}\right)}}{\sqrt{\sum_k{{\left(s_{u,k}-\overline{s_u}\right)}^2\sum_k{{\left(s_{v,k}-\overline{s_v}\right)}^2}}}}$} & Where $s_{u,k}$ is the position of $P_{u,k}$ in the decreasing order ranking of Big Five scores. \\ \hline 
	\cite{FAFinder2020} & Hellinger-Bhattacharya Distance & \resizebox{0.91\hsize}{!}{$SimP\left(u,\ v\right)=\frac{1}{\sqrt{2}}*\sqrt{\sum^5_{i=1}{{\left(\sqrt{u_i}-\sqrt{v_i}\right)}^2}}$} &  \\ \hline 
\end{tabular}
\end{table*}

\subsubsection{Matrix factorization methods}
In personality-enhanced matrix factorization methods, the conventional matrix factorization algorithm is extended to incorporate the user's personality traits along with its ratings. In the conventional matrix factorization method, the user-item interaction matrix is decomposed to the product of two low-dimensionality rectangular matrices that represent the represent users and items in a lower dimensional latent space, this is done by applying dimensionality reduction algorithm such as singular value decomposition. Formally, let $p_u\in {\mathbb{R}}^k$ and $q_i\in {\mathbb{R}}^k$ denote the latent feature vector of user $u$ and item $i$ respectively. In the conventional matrix factorization method, the user u's preference to item $i$ is estimated by computing the dot product of user $u$ and item $i$ latent feature vectors, as shown in eq(\ref{equ_rating1}).

\begin{equation}
\label{equ_rating1}
{\hat{r}}_{ui}=p_u.q_i
\end{equation}

Personality-enhanced matrix factorization extends this by incorporating the user's personality traits, and other occasionally other social attributes. They introduce an additional latent feature vector $y_a\in {\mathbb{R}}^k$ for each social attribute $a\in A$. eq(\ref{equ_rating1}) is extended to incorporate these attributes as shown in eq(\ref{equ_rating2}). It is worth noting that some works consider the Big-Five personality score vector as the only attribute, as in \cite{Fernandez_Tobias2016}. While some other works add other social attributes in addition to the personality, for instance, in \cite{Elahi2013} the user's gender, age group and the scores for the Big Five personality traits are used as attributes. 

\begin{equation}
\label{equ_rating2}
{\hat{r}}_{ui}=q_i.\left(p_u+\sum_{a\in A}{y_a}\right)
\end{equation}

\subsection{Personality matching methods}
This approach is similar to the conventional content filtering approach. In the personality matching method, a personality score is assigned to each item. This is done either using content analysis, attribute analysis or a hybrid approach. In the content analysis, the system assigns a personality score to an item by applying an automatic personality recognition on the content of that item, such as the textual description, labels and category. For example, in \cite{Buettner2017} a product personality assessment method known as product personality scale \cite{Mugge2009} was used to assess the personality of the items. While in attribute analysis, the system assigns a personality score to an item by analyzing the attribute of that item, such as the personality traits of users that interacted with that item. For instance, in \cite{Yang2019} the personality $P_{G_i}$ of a video game $G_i$ is assigned by computing the average personality traits of users who played $G_i$ as $P_{G_i}=\frac{\sum_{U_j\in O_{G_i}}{P_{U_j}}}{\left|O_{G_i}\right|}$. Personality matching is usually applied if we can observe a common matching criteria between the recommended content and the target user, which eliminate the need for extensive computing in order to find the neighborhood set from one hand, and mitigate the cold start on the other hand. Xiao \textit{et al.} \cite{Xiao2018} used personality matching approach for followee recommendation, the total personality matching (TPM) score between a given user $u$ and the potential blogger $pf$ is computed as shown in eq(\ref{equ_TPM}):

\begin{equation}
\label{equ_TPM}
TPM(u,\ pf)\ =\ \mu (\ MS(u,\ pf,\ dim))
\end{equation}

Where $MS(u,\ pf,\ dim)$ denotes the personality matching score of the user $u$ and the potential recommendation followee $pf$ in a the respective dimension, and $\mu $ is the average value of each dimension. 

\subsection{Hybrid personality filtering methods}
Hybrid personality filtering methods combine the technique of personality filtering on the users' space, and personality matching on the items space. Hybrid personality filtering has been proven as an effective method that leverages the advantages of personality filtering and personality matching methods. \cite{ning2019personet,Dhelim2020entity} used a hybrid personality filtering approach for friend recommendation, where personality filtering is used to find users with similar ratings and personality matching is used to filter the item space (since it is a friend recommendation system, the items represent potential friends). Similarly, \cite{Yang2019} also used hybrid personality filtering for a game recommendation, where personality filtering is used to determine user with similar game ratings, and personality matching is used to attribute personality to games.

\begin{table}[]
	\centering
	\caption{Personality-aware recommendation techniques classification}
	\label{schemes_classification}
	\begin{tabular}{|p{1.25in}|p{1.0in}|} \hline 
		\textbf{Recommendation method} & \textbf{Publication}  \\ \hline 
		Personality neighborhood & \cite{Wu2015,Zhou2011,Hariadi2017,Asabere2020,ning2019personet,Xia2017,Asabere2018,YusefiHafshejani2018,Hu2011,Wu2018,Buettner2017,Yang2019,Wu2015,Fernandez_Tobias2014,FAFinder2020} \cite{fernandez2015,Wang2015,hu2010using}   \\ \hline 
		Matrix factorization & \cite{Elahi2013,Fernandez_Tobias2016,Aguiar2020,braunhofer2014,Feng2013,Zheng2020,sun2018,potash2016recommender,Khodabandehlou2020}   \\ \hline 
		Personality  matching & \cite{Buettner2017,LiBian2012,cheng2016,Tanasescu2013} \cite{Xiao2018}   \\ \hline 
		Hybrid personality filtering & \cite{ning2019personet,Dhelim2020KBS} \cite{Yang2019,Dhelim2020TCSS,Dhelim2020IOT}   \\ \hline 
		Deep-learning & \cite{Moscato2020,Majumder2017,Mikolov2013,wei2018,jeong2020adaptive,Neehal2019,He2020,Li2019,Li2020,jeong2020adaptive,Zhang2018}   \\ \hline 
	\end{tabular}
\end{table}

\subsection{Deep learning methods}
In recent years, deep learning has revolutionized the domain of recommendation systems by leveraging deep learning models, such as Convolutional Neural Network (CNN), Recurrent Neural Network (RNN) and Autoencoder, to name a few. Personality-aware recommendation systems are not an exception to this revolution. Deep learning is either used to detect the user personality of the users or in the recommendation process itself. The choice of deep learning model used for personality detection and personality-aware recommendation depends on the type of source data of users. Deep learning model that are inspired by natural language processing, such as the n-gram model, are suitable for personality detection and content recommendation from textual source data. For instance, Majumder \textit{et al.} \cite{Majumder2017} proposed deep CNN for document-level automatic personality recognition, the CNN extracts monogram, bigram and trigram features from the document text and each word was represented in the input as a fixed-length feature vector using Word2Vec \cite{Mikolov2013} model, finally linguistic features (e.g. LIWC, MRC) are concentrated and fed to fully connected layer for personality traits prediction. Similarly, deep learning models that are designed for image and video processing are suitable for personality detection and personality-aware recommendation using non-textual personality data. Wei \textit{et al.} \cite{wei2018} proposed Deep Bimodal Regression (DBR) framework for apparent personality analysis from videos and images. DBR modified the common convolutional neural networks for incorporating essential visual cues.  Besides the source data format, the recommended content nature could also influence the choice of deep learning models to be used as personality-aware recommendations. For example, Chi-Seo \textit{et al.} \cite{jeong2020adaptive} introduced a system that employs deep learning to classify and recommend tourism types that are compatible with the user's personality. The model is composed out of three layers, each layer incorporates a service provisioning layer that real users face, the recommendation service layer, responsible to produce recommended services based on user information inputted, and the adaptive definition layer, that learns the types of tourism that fits for the user's personality types.

In Table \ref{schemes_classification} we classify some of the recent personality-aware recommendation systems based on their recommendation technique.

\section{Literature review of personality-aware recommendation systems}
In the last few years, we have witnessed a rapid proliferation of personality-aware recommendation systems. In this section, we review the literature on personality-aware recommendation systems in different application domains.
\subsection{Friend recommendations}
In the literature of social networks, many personality-aware friend recommendation systems have been proposed, Ning \textit{et al.} \cite{ning2019personet} proposed a personality-aware friend recommendation system named PersoNet that leverages Big-Five personality traits to enhance the hybrid filtering friend selection process. PersoNet outperformed the conventional rating-based hybrid filtering, and achieve acceptable precision and recall values in cold start phase as well. Similarly, Chakrabarty \textit{et al.} \cite{FAFinder2020} designed a personality-aware friend recommendation system name FAFinder (Friend Affinity Finder). FAFinder uses Hellinger-Bhattacharyya Distance (H-B Distance) to measure the user's Big-Five similarity and recommend friends accordingly. While Bian \textit{et al.} \cite{LiBian2012} designed and implemented Matchmaker, a personality-aware friend recommendation system that recommends friends to users on Facebook by matching and comparing user's online profile with the profiles of TV characters. For example, if Facebook user X is similar to TV character 1, and Facebook user Y is similar to TV character 2, and character 1 and character 2 are friends in the same TV show, then the Matchmaker system recommends user X to become friends with user Y. Whereas, Neehal \textit{et al.} \cite{Neehal2019} introduced a personality-aware friend recommendation framework, which uses a 3-Layered artificial neural network (ANN) for friend preference classification and a distance-based sorted subset selection function for friend recommendation.

Tommasel \textit{et al.} \cite{Tommasel2016,tommasel2015role} studied the effects of user personality on the accuracy of followees prediction on microblogging social media. The authors analysed how the user's personality character influence the followees selection process by incorporating personality traits with state of the art followee predicting factors. To prove the effectiveness of proposed followee prediction algorithm, the author collected a Twitter dataset by crawling the account of 1852 users, and only users that English is their tweeting language were selected. They tested the content-based followee prediction algorithm with and without including the user's personality traits. Their results showed that incorporating personality traits can enhance the followee recommendations. While in Tommasel \textit{et al.} \cite{Tommasel2015}, they analyzed 3 different similarity factors. Firstly, they calculated the total similarity by taking into account the Big Five personality factors as a whole. Secondly, they calculated the dimension to dimension similarity measure by taking into account every individual personality traits separated from each other. Finally, they calculated a cross dimension similarity measurement by taking into account every personality faced in relation to the others. Their results showed that personality traits must be regarded as a distinctive factor in the process of followee prediction. However, personality dimensions should not be analyzed as a whole because the overall personality similarity measurement might not precisely assess the actual matching between users. The data analysis proves the existence of relations among the individual personality facades. Therefore, the importance of assessing each personality trait with respect to other users. Similarly, Xiao \textit{et al.} \cite{Xiao2018} introduced a personality-aware followee recommendation system based on sentiment analysis and text semantics, they proposed model combines the user attributes with the Big-Five traits to recommend new followees. And Mukta \textit{et al.} \cite{Mukta2016} proposed a technique to detect homophily by analyzing the Big-Five personality traits of users in an egocentric network such as Facebook.

\subsection{Movies recommendations}
Asabere \textit{et al.} \cite{Asabere2020} proposed ROPPSA, a personality-aware TV program recommendation system that leverages normalization and folksonomy procedures to generate group recommendations for viewers with similar personality traits. Balakrishnan \textit{et al.} \cite{Balakrishnan2018} proposed a hybrid recommender system for movies named HyPeRM,, the proposed system includes the users' personality character in addition to their demographic information (e.g. sex and age) to enhance the precision of the recommendations. Big Five personality trait was employed to measure the users' personalities. HyPeRM was tested based on the Root Mean Square Error of Approximation (RMSEA) and the Standardized Root Mean Square Residual (SRMR). Both theses metrics showed that HyPeRM outperformed the baseline variant (i.e. the recommendation without including the user's personality) in terms of the precision of the recommendation. Their work shows that movies recommendations can be improved by incorporating the viewers' personality traits. Similarly, Shanchez \textit{et al.} \cite{QuijanoSanchez2011} proposed HappyMovie, a Facebook application for movie recommendations, HappyMovie uses three features for a movie recommendation, user personality, social trust with other users and the past movie ratings. While Bolock \textit{et al.} \cite{bolock2020} proposed a movie recommendation system based on the user's character. The system is adaptive in the way it uses a different recommendation algorithm for different users based on the used character criteria. The authors implemented a movie recommendation application to find the relationship between the user's character and the recommendation algorithm, they have used three main character dimensions, user personality, background and gender. 

On the relationship between personality and movie preferences. Golbeck \textit{et al.} \cite{Golbeck2013} proved the positive correlation between personality and users' movie preferences. Using surveys and analysis of system data for 73 Netflix users, they proved correlations between personality and preferences for specific movie genres. Wu and Chen \cite{Wu2015} studied user personality inferences using implicit behaviors with movies, and the possibility to recommend movies base on the user's personality traits without users' explicit ratings. Specifically, they determined a set of behavioral features using experimental confirmation and proposed an inference method using Gaussian Process to fuse these features and subsequently detect the user's Big-Five personality traits. After that, they used the obtained personality information to enhance the collaborative filtering movie recommendation process. Scott \textit{et al.} \cite{Scott2016} investigated the relationship between personality and cultural traits with a perception of multimedia quality. Hu \textit{et al.} \cite{Hu2011} addressed the cold-start problem by incorporating human personality into the collaborative filtering framework, they have tested the proposed system with movies and music public datasets. And Berkovsky \textit{et al.} \cite{berkovsky2017recommend}, studied the effects of different recommendation and content filtering strategies on user trust. They evaluated the score of nine main factors of trust grouped and divided them into three dimensions and tested the different observations regarding the users' personality traits. 

\subsection{Music recommendations}
The authors of the works presented in \cite{klec2017influence,ferwerda2017personality,ferwerda2016personality} discussed the impact of personality traits on the accuracy of music recommendation systems. Cheng \textit{et al.} \cite{cheng2016} Introduced a hybrid method for personality-aware music recommendations. They used personality matching of the user's personality traits with an extracted feature from songs audio, and classify these features using support vector machine (SVM) algorithm. While Schedl \textit{et al.} \cite{Schedl2016} studied the relationship between personality traits and classical music preferences, they grouped the users into four clusters based on the personality traits and tried to infer the preference of each cluster regarding classical music. Ferwerda \textit{et al.} \cite{ferwerda2014enhancing} discussed the possibility to enhance music recommendation systems by incorporating the user's psychological factors such as emotional and personality states. The authors discussed how people listen to music to control their emotional states, and how this adjustment is related to their personality traits. They focused on the methods to acquire data from social media networks to estimate the current emotional state of the listeners. Finally, they discussed the connection of the accurate emotionally with the music categories to support the emotional adjustment of listeners. The same research group proposed a personality-aware music recommender system \cite{ferwerda2016personality}, where they employed the users' personality traits as a general model. The authors specified the relationships between listeners' personality and their behavior, preferences, and needs, in addition to that the authors studied the different ways to infer users' personality traits from user-generated data of social media websites (e.g., Facebook, Twitter, and Instagram). Hu \textit{et al.} \cite{Hu2010a} proposed a general model that can deduce users' music preferences based on their personality characteristics. Their subject studies prove that most of the active users think that the recommended songs are more precise for their friends, however, these users enjoy more using personality questionnaires based recommenders for finding songs for themselves. The authors investigate if domain based knowledge has an impact on users' understanding of the system. We found that novice users, who are less knowledgeable about music, generally appreciated more personality based recommenders. Zhou \textit{et al.} \cite{Zhou2011} used decision trees to developed a heuristic personality-aware music recommendation system for niche market.

\subsection{Image recommendation}
Associating images features and personality traits is twofold: known the feature of the image can help to infer the personality of users who interact with the image, and know the personality traits of the users could help to recommend relevant images. Guntuku \textit{et al.} \cite{guntuku2015} studied methods for modeling the personality character of users based on a collection of images that they tagged as `favorite' on Flickr. Their study presents several methods for enhancing personality detection performance by proposing better features and modeling approaches. They evaluated their approach by measuring its efficiency when used in an image recommendation system. The presented results showed the need for using high-level user understandable features and illustrate the effectiveness of a +A2P (Answers-to-Personality) and F2A (Features-to-Answers) approaches compared to the traditional F2P (Features-to-Personality) method that is usually used by existing works. While in \cite{Guntuku2015enjoy} they studied the effects of personality (Big-Five Model) and cultural traits (Hofstede Model) on the potency of multimedia-stimulated positive and negative emotions. Wu \textit{et al.} \cite{Wu2019} proposed a hierarchical attention model for social feature based image recommendation, in which the recommendation system considers the social characteristics of the users, such as user social interests and personality traits. Li \textit{et al.} \cite{Li2019} developed a deep-learning-based image aesthetic model that employs the user's personality traits for image aesthetic rating. The personality features are used to represent the aesthetics features, hence, producing the optimal generic image aesthetics scores. Furthermore, in \cite{Li2020} they extended their method to offer a personality-aware multi-task framework for generic as well as personalized image aesthetics assessment. Gelli \textit{et al.} \cite{Gelli2017} investigated the effects of personality on user behaviors with images in a social media, and which visual stimuli contained in photo content can affect user behaviors. They analyzed a twitter dataset of 1.6 million user and image retweet behaviors. Kim \textit{et al.} \cite{Kim2019} studied the relationships between Instagram user personality traits and color features of their photos, and found that agreeableness is the most relevant trait that is associated with the photo and color features.

\subsection{Academic content recommendations}
Many works have used personality traits for academic-oriented recommendation systems, such as courses recommendations, conference attendee recommendations and research paper recommendations. Xie el al \cite{Xia2017} proposed a recommendation system of academic conference participants called SPARP (Socially-Personality-Aware-Recommendation-of-Participants). For more effective collaborations in the vision of a smart conference, the proposed recommendation approach uses a hybrid model of interpersonal relationships among academic conference participants and their personality traits. At first, the proposed system determines the social ties among the participants based on past and present social ties from the dataset with four trial-weight parameters. These weight parameters are used later in their experiment to represent various influence proportions of the past and present social ties among participants. Following that, the system calculates the personality-similarity between the conference participants based on explicit tagged-data of the personality ratings. Similarly, Asabere \textit{et al.} \cite{Asabere2018} proposed a recommendation algorithm for conference attendees called PerSAR (Personality-Socially-Aware-Recommender). The proposed system is based on a hybrid approach of social relations and personality characters of the conference participants. To evaluate their proposed system, the authors used the dataset of The International Conference on Web-Based Learning (ICWL) 2012, which includes the social ties of 78 conference participants with a total time-frame of 12 hours (720 minutes). Far from that, Fahim Uddin \textit{et al.} \cite{Uddin2016} Proposed a personality-aware framework to improve academic choice for newly enrolled students. Their proposed framework makes use of the research field of Predicting Educational Relevance For an Efficient Classification of Talent, which uses stochastic probability distribution modeling to help the student to choose the relevant academic field. Hariadi et el \cite{Hariadi2017} proposed a personality-aware book recommendation system that combines the user's attributes as well as his personality traits. The proposed system leverages MSV-MSL (Most Similar Visited Material to the Most Similar Learner) method to compute the similarity between users and form the personality neighborhood.

\subsection{Product recommendations}
Dhelim \textit{et al.} \cite{Dhelim2020TCSS} introduced Meta-Interest, a personality-aware product recommendation system that considers the user interests and personality traits and recommends relevant products by exploring the possible user-item metapaths. Tkalcic \textit{et al.}~\cite{Marko2010} proposed a new approach for measuring the~user similarity~for collaborative filtering~recommender systems that is~based~on the Big Five~personality~model in the context of product recommendation. Buettner \cite{Buettner2017} introduced a personality-aware framework for product recommender named PBPR. The proposed framework analyzes the user's social media profile to infer its personality traits and recommend products accordingly. The author evaluated his proposed framework as IT artefact using a dataset from XING. Huang \textit{et al.} \cite{Huang2020}, used a data driven method to predict online shoppers' online buying preferences. Firstly, the authors used text mining method based on the shoppers' language usage behaviors to create seven different dimension lifestyle-lexicons. Following that, they included these lifestyle-lexicons in the product recommendation system that can predict the shoppers' buying preferences. Roffo \cite{roffo2016towards} discussed utilizing personality to compute the association between the shopper's purchasing tendency and the advert's recommendations. Moreover, the author introduced the ADS dataset, an advertising benchmark enriched with Big Five personality traits of users along with 1200 personal photos. Adamopoulos and Todri \cite{adamopoulos2015personality} used a dataset from Amazon.com to evaluate a personality based recommendation, they have inferred the users' personality traits along with their needs and other contextual information from their social media profiles. Their findings is that adding personality to the recommendation process can increase the efficiency of the system. 
\subsection{Game recommendations}
Yang \textit{et al.} \cite{Yang2019} introduced a personality-aware game recommendation system, they apply text mining on the players' social network posts to extract their personality types and analyzed the games' content to associate these games with certain personality types. They proved the effectiveness of their proposed system through an experiment on 63 players and more than 2000 games. Lima \textit{et al.} \cite{DeLima2018} designed a new method for interactive storytelling in games, in which the quests and the ongoing story follow the view of individual personality traits and behaviors in a non-deterministic way. Chan \textit{et al.} \cite{Chan2018} proposed a method for matching players using personality types to augment the enjoyment and social interaction in exergames. They argue that maintaining high levels of enjoyment and active social interactions is crucial because both can offer retention and continuation of gameplay and exercise involvement. Hill \textit{et al.} \cite{Zeigler_Hill2015} investigated the association between HEXACO personality model with preferences for certain aspects of gaming experiences. The main finding confirmed that extraversion trait is moderately associated with the socializer gaming preference and a slight association with the daredevil gaming preference. While Abbasi \textit{et al.} \cite{Abbasi2020} discussed the personality differences between gamers and non-gamers. Supported by evidence obtained by analyzing the personality types of 855 students (gamers and non-gamers), they conclude that gamers have a personality types that is significantly different on compared to non-gamers.

\subsection{Points of interest recommendations}
Wang \textit{et al.} \cite{Wang2020} proposed a trust-based POI recommendation system, they leverage the personality similarity between users to compute the trust level. In addition to trust and personality information, they also make use of the graphic and temporal influence in the recommendation model. Chi-Seo \textit{et al.} \cite{jeong2020adaptive} introduced a system that employs deep learning to classify and recommend tourism types that are compatible with the user's personality. The model is composed out of three layers, each layer incorporates a service provisioning layer that real users face, the recommendation service layer, responsible to produce recommended services based on user information inputted, and the adaptive definition layer, that learns the types of tourism that fit for the user's personality types. Zhang \textit{et al.} \cite{Zhang2018} introduced a new POI recommendation system that uses POI classification model named POIC-ELM. POIC-ELM extracts 9 features that are related to 3 factors, the user's personality information the POI information and the user's social relationships information. The learned feature are then fed to an extreme learning machine (ELM) for POI classification. Braunhofer \textit{et al.} \cite{braunhofer2014} introduced STS (South Tyrol Suggests), a personality-aware POI recommender system that uses an active learning module and personality-aware matrix factorization recommendation to infer the relevant POI. In the same vein, in \cite{braunhofer2014context} they designed a personalized active learning method that takes advantage of the user's personality information to get more accurate in-context POI ratings. Tanasescu \textit{et al.} \cite{Tanasescu2013} introduced the concept of 'personality of a venue'. They extracted keywords and other annotations from the reviews of the venues and mapped these information to Big-Five personality traits. The experimental testing confirmed the correlation between visitors' personality traits and the personality of the visited venue. Sertkan \textit{et al.} \cite{Sertkan2019} proposed an automatic method for computing the Seven-Factor equivalent of tourism sites. Regression analysis, cluster analysis, and exploratory data analysis are performed to find the correlation between Seven-Factors and the type of tourist destination site. Feng \textit{et al.} \cite{Feng2013} fused three factors, mainly interpersonal interest similarity, personal interest similarity and interpersonal influence to implement probabilistic matrix factorization for personality-aware recommendations.

In Table \ref{content_classification}, we summarize the reviewed works related to personality-aware recommendation systems based on the recommended content.

\begin{table}[]
	\centering
	\caption{Personality-aware personality recommendation system content-based classification}
	\label{content_classification}
\begin{tabular}{|p{1.2in}|p{1.5in}|} \hline 
	\textbf{Recommendation domain} & \textbf{Recommendation system} \\ \hline 
	Friend recommendation & \cite{ning2019personet} \cite{Mukta2016,FAFinder2020,LiBian2012,Neehal2019,Tommasel2016,tommasel2015role,Tommasel2015,Xiao2018,bian2011online} \\ \hline 
	Image recommendation & \cite{guntuku2015,He2020,Guntuku2015enjoy,Wu2019,Li2019,Li2020,Gelli2017,Kim2019,Marko2010} \\ \hline 
	Movie recommendation & \cite{berkovsky2017recommend,Balakrishnan2018,Wu2015,QuijanoSanchez2011,bolock2020,Golbeck2013,Karumur2016,potash2016recommender,Scott2016,Hu2011,5050recommend,Recio_Garcia2009,Quijano_Sanchez2010,sofia2016investigating,odic2013personality,Wu2013,tkalcic2014personality,Khan2020,Dhelim2018} \\ \hline 
	Music recommendation & \cite{Moscato2020,Gupta2020,Schedl2016,Zhou2011,cheng2016,klec2017influence,ferwerda2017personality,ferwerda2016personality,ferwerda2014enhancing,Hu2010a,ferwerda2016influence,onori2016comparative,Ferwerda2017,Hu2010,Ferwerda2015,Melchiorre2020a,Bansal2020,Kouki2020,Liu2020,Millecamp2020} \\ \hline 
	Product recommendation & \cite{Buettner2017,Aguiar2020,roffo2016towards,Marko2010,Huang2020,adamopoulos2015personality,enlighten149660,Hu2014,hu2010using,Khodabandehlou2020} \\ \hline 
	POI recommendation & \cite{jeong2020adaptive,braunhofer2014,Feng2013,Zhang2018,Wang2020,Sertkan2019,Tanasescu2013,braunhofer2014context,Braunhofer2015,Alves2020,laban2020don,Khelloufi2020,Cai2020} \\ \hline 
	Academic recommendation & \cite{Zheng2020,Xia2017,Asabere2018,Uddin2016,Hariadi2017,Uddin2016,celli2018big,Qamhieh2020} \\ \hline 
	Game recommendation & \cite{Yang2019,DeLima2018,Chan2018,Zeigler_Hill2015,Abbasi2020,Yang2017} \\ \hline 
\end{tabular}
\end{table}

\section{Datasets and Benchmarks}
Due to the availability of open public datasets that considered the users' personality information, many personality-aware recommendation systems were able to train their proposed models and compare them using the state-of-the-art benchmarks. In this section, we present two of the widely used personality datasets in the context of personality-aware recommendation systems.

\subsection{myPersonality dataset}
In 2007, David Stillwell a PhD student at the University of Nottingham designed a Facebook application called myPersonality that leverages IPIP version of the NEO personality inventory personality questionnaire and displays the personality score instantly \cite{stillwell2004mypersonality}. myPersonality was initially intended for limited use, David shared it with his close friends. Later on, surprisingly the number of users who joined the study increase dramatically, and many users were willing to donate their data to be used for academic purposes. By 2012, more than 6 million users finished the IPIP personality questionnaire, and the respondents came from different age groups, backgrounds and cultures. myPersonality dataset was anonymized and samples of it were shared with many researchers. In 2018, the creators of myPersonality decided to stop the project, as it has become extremely challenging to maintain the dataset with the increasing number of usage requests from researchers over the last few years.
\subsection{MovieLens dataset}
MovieLens is a widely used open dataset in recommendation system researches. It contains movie rating data extracted from the famous movie recommendation and rating website MovieLens.com \cite{Harper2016}. The ratings were collected over different periods of time, there are many available versions of the dataset depending on the size of the dataset. The largest available version of the dataset is named MovieLens 25M. It contains 25 million movie ratings and one million tag applications applied to 62,000 movies by 162,000 users. Personality2018 \cite{Nguyen2018} is a version of MovieLens dataset that includes the personality information of the users that rated the movies, it contains the TIPI score of 1834 users along with the movie rating that were given by these users. 

\subsection{Newsfullness dataset}
Newsfullness is a news sharing platform that uses personality-aware recommendation for of news articles \cite{Dhelim2020KBS}. Newsfullness contains more the TIPI score of 2228 users along with their articles that these users viewed or liked. The collected articles were from all the main news websites, such as BBC, CNN, RA and Aljazeera, from different news categories (business, politics, health, sports, travel, entertainment, art, education, science and technology). Table \ref{datasets} summarizes the works that have used these datasets.

\begin{table}[]
	\centering
	\caption{Personality datasets}
	\label{datasets}
\begin{tabular}{|p{0.7in}|p{1.8in}|p{0.8in}|} \hline 
	\textbf{Dataset} & \textbf{Works} & \textbf{Content} \\ \hline 
	myPersonality & \cite{sun2018,fernandez2015,cantador2013relating,onori2016comparative,potash2016recommender,Ferwerda2017,celli2018big,Moscato2020} & N/A \\ \hline 
	MovieLens & \cite{Karumur2016a,5050recommend,Recio_Garcia2009,Wang2015,Karumur2016,Hu2011,Hu2010,hu2010using,Zhou2011} & Movies \\ \hline 
	Newsfullness & \cite{Dhelim2020KBS,Dhelim2020TCSS,Dhelim2020IOT} & News articles \\ \hline 
	Twitter & \cite{Khan2020,celli2018big,Tommasel2016,Tommasel2015,tommasel2015role} & Friendships \\ \hline 
	IMDB & \cite{bian2011online,berkovsky2017recommend,Khan2020} & Movies \\ \hline 
	Last.fm & \cite{onori2016comparative,Ferwerda2017,Melchiorre2020a,Kouki2020} & Music \\ \hline 
	Other datasets & allrecipes.com \cite{adaji2018}, Amazon\cite{Yakhchi2020}, Steam \cite{Yang2017,Yang2019} ADS \cite{roffo2016towards,enlighten149660}, Douban \cite{Wu2015,Wu2013,Wu2017}, IAPS\cite{Marko2010}, Facebook \cite{Mukta2016}, , PsychoFlickr \cite{guntuku2015}, Sina Weibo\cite{Xiao2018}, Deezer \cite{Moscato2020} & N/A \\ \hline 
\end{tabular}
\end{table}

\section{Challenges and open issues}
Although that personality-aware recommendation system offers many advantages and solutions to tackle recommendation challenges that conventional recommendation systems cannot solve, such as cold start and recommendation diversity. However, using the user's personality in the recommendation bring up new challenges and ethical issues, in this section we discuss some of these challenges.
\subsection{Personality information privacy}
The privacy of the user's personality poses a new challenge in addition to the existing challenge of preserving the privacy of user's information. As the user's personality information is even more sensitive than other information in the user's profile. In March 2018, Facebook-Cambridge Analytica scandal has drawn the attention of the world. A Facebook application created by the a data analytic company named Cambridge Analytica unrightfully collected the personality information of more than 87 million users, aiming to manipulate their voting choice in the 2016 US presidential election \cite{Hinds2020}. The challenge of personality-aware recommendation system is to preserve the personality information of the users, as malicious users can analyze the recommendation results to predict the personality type of other users. The recommendation system is responsible to maintain the tradeoff between data sharing and information privacy between users.

\subsection{Measurement accuracy}
The accuracy of the personality measurement is vital for personality-aware recommendation system, the inaccurate measurement of the user personality traits will inevitably lead to inaccurate recommendations. What makes things worse is that the system considers personality traits as content information that do not need update frequently, and will offer inaccurate information all the time. The personality questionnaire contains questions that are relative to the subject itself, and there is no standard measurement of the questioned features, which could increase the reference-group effect. For instance, an introverted subject may identify himself as an extrovert, even if he filled the questionnaire correctly, that is because all his close friends are also introverts, therefore his judgment was relative to his environment. APR methods may also inaccurately detect the user's personality for various reasons, for example, image-based APR might predict the personality of a user by analyzing his shared photos on social media without considering the context of these photos. For example, a user who shares nature photos frequently as a part of his job as a photographer or a war photo shared by a journalist may not reflect their personalities.

\section{Conclusion}
To the best of our knowledge, this survey is the first that focuses on personality-aware recommendation system. We have reviewed the literature of the recent works in this domain, and show the main differences between different works, in terms of personality model, as well as in terms of the used recommendation technique. The vast majority of personality-aware recommendation systems leverage Big-Five personality model to represent the user's personality. Personality-aware recommendation systems have the upper hand when compared with the conventional recommendation techniques, especially when dealing with cold start and data sparsity problems. However, with the understanding of the user's personality advantage comes the challenge of preserving the privacy of the user personality information, and also the challenge of maintaining a high personality detection accuracy.

\section*{Acknowledgments}
We would like to thank all the active users of Newsfullness that agreed to be a part of the Meta-Interest experiment. 
This work was supported by the National Natural Science Foundation of China under Grant 61872038, and in part by the Fundamental Research Funds for the Central Universities under Grant FRF-BD-18-016A.

\ifCLASSOPTIONcaptionsoff
  \newpage
\fi



%



\bibliographystyle{IEEEtran}
\bibliography{IEEEabrv,refs}

%

\vskip 0pt plus -1fil

\begin{IEEEbiography}[{\includegraphics[width=1in,height=1.25in,clip,keepaspectratio]{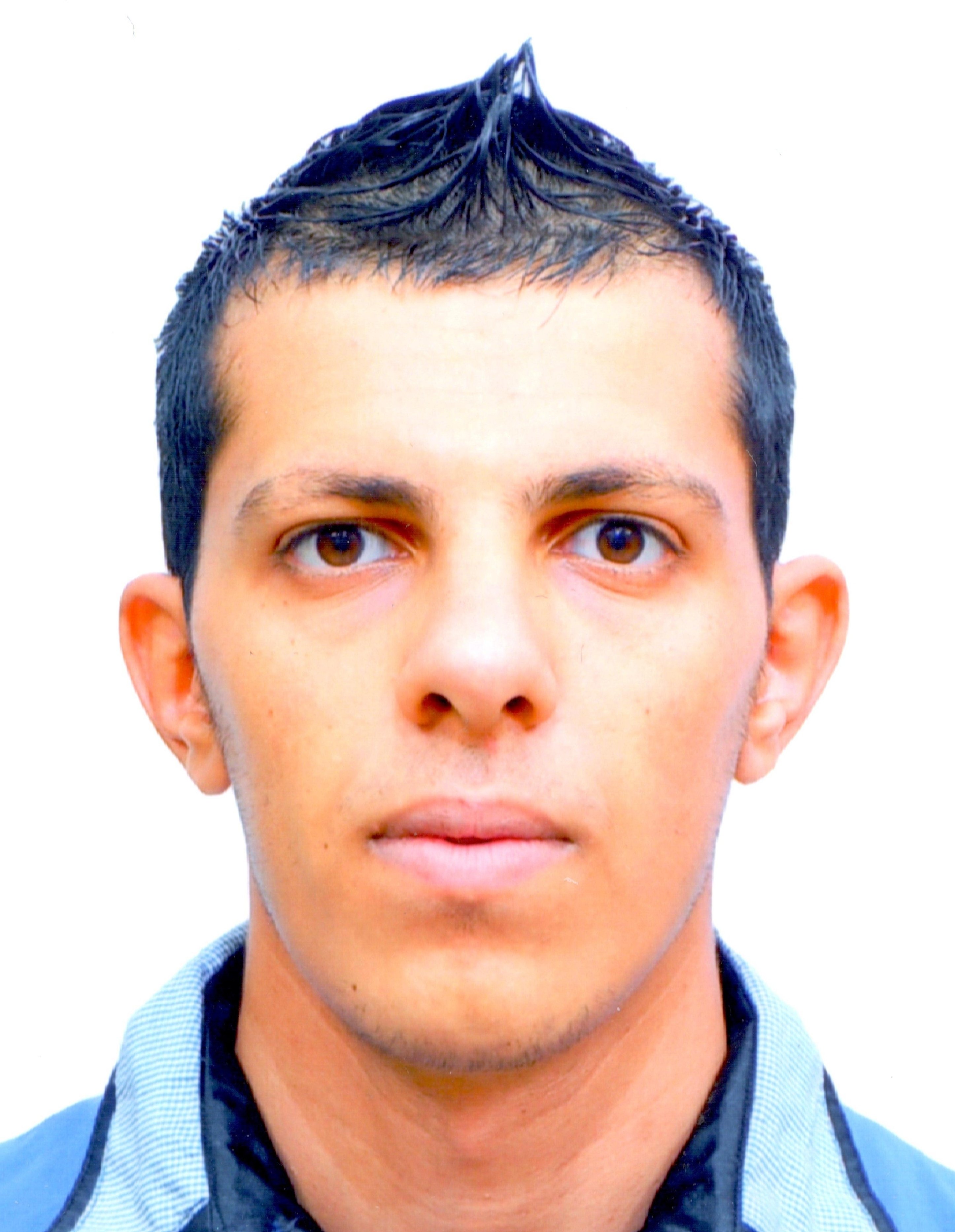}}]{Sahraoui Dhelim}
	Received his B.S. in Computer Science from the University of Djelfa, Algeria, in 2012 and his Master degree in Networking and Distributed Systems from the University of Laghouat, Algeria, in 2014. And PhD in Computer Science and Technology from University of Science and Technology Beijing, China, in 2020. His current research interests include Social Computing, Personality Computing, User Modeling, Interest Mining, Recommendation Systems and Intelligent Transportation Systems.
\end{IEEEbiography}

\vskip 0pt plus -1fil

\begin{IEEEbiography}[{\includegraphics[width=1in,height=1.25in,clip,keepaspectratio]{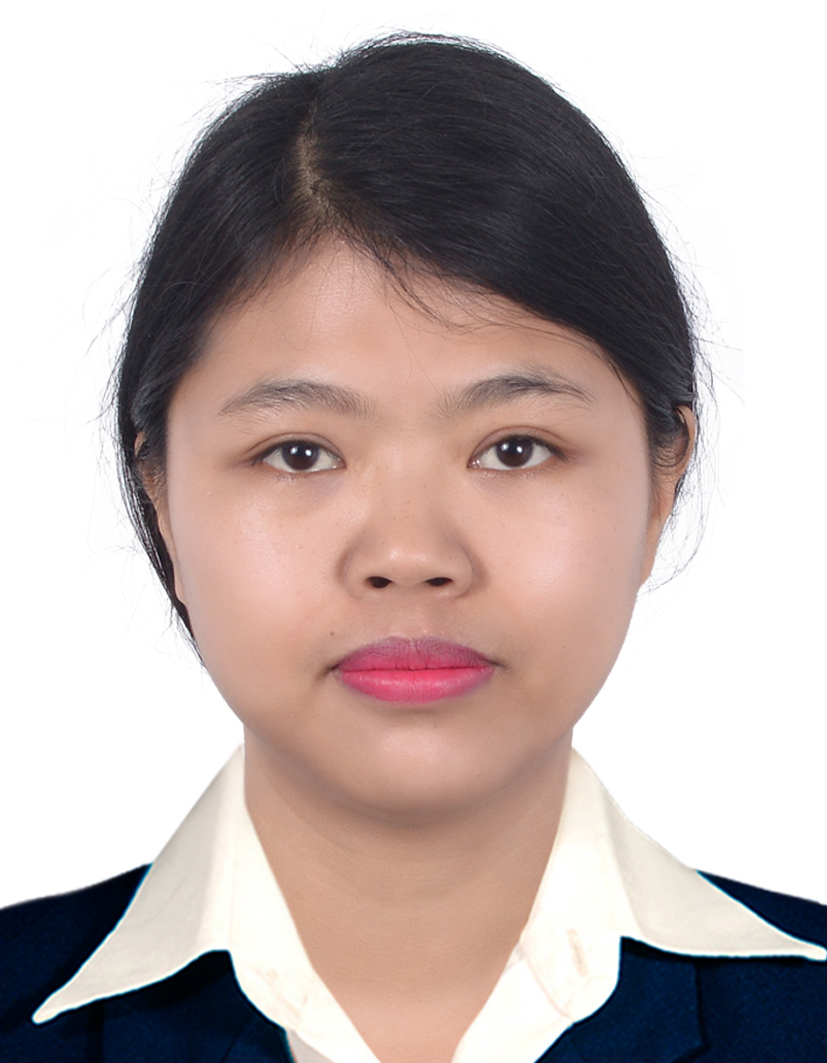}}]{Nyothiri Aung}
	Received Master of Engineering (Information Technology) degree form Mandalay Technological University, Myanmar, 2012. And PhD in Computer Science and Technology from University of Science and Technology Beijing, China, 2020. She worked as a tutors at the Department of Information Technology in Technological University of Meiktila, Myanmar (2008-2010). And System Analyst of ACE Data System, Myanmar (2012-2015). Her research interests include Social Computing, Personality Computing and Intelligent Transportation System.
\end{IEEEbiography}

\vskip 0pt plus -1fil
\begin{IEEEbiography}[{\includegraphics[width=1in,height=1.25in,clip,keepaspectratio]{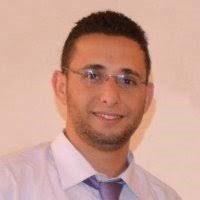}}]{Mohammed Amine Bouras}
	Received B.S. and M.S degree from University of Laghouat, Algeria, the school of computer science in 2015. He is currently a Ph.D candidate in University of science and Technology Beijing, China, school of computer and communication Engineering. He focuses on the convergence in the Internet of Things and Big Data. His research interest includes Internet of Things, semantic internet of things, Big Data, Data analysis, Cyber Physical Social Thinking (CPST) Spaces.
\end{IEEEbiography}

\vskip 0pt plus -1fil

\begin{IEEEbiography}[{\includegraphics[width=1in,height=1.25in,clip,keepaspectratio]{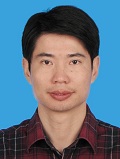}}]{Huansheng Ning}
	Received his B.S. degree from Anhui University in 1996 and his Ph.D. degree from Beihang University in 2001. Now, he is a professor and vice dean of the School of Computer and Communication Engineering, University of Science and Technology Beijing, China. His current research focuses on the Internet of Things and general cyberspace.
	He is the founder and chair of the Cyberspace and Cybermatics International Science and Technology Cooperation Base.
	He has presided many research projects including Natural Science Foundation of China, National High Technology Research and Development Program of China (863 Project). He has published more than 100+ journal/conference papers, and authored 5 books. He serves as an associate editor of IEEE Systems Journal (2013-Now), IEEE Internet of Things Journal (2014-2018), and as steering committee member of IEEE Internet of Things Journal (2016-Now).
\end{IEEEbiography}

\vskip 0pt plus -1fil

\begin{IEEEbiography}[{\includegraphics[width=1in,height=1.25in,clip,keepaspectratio]{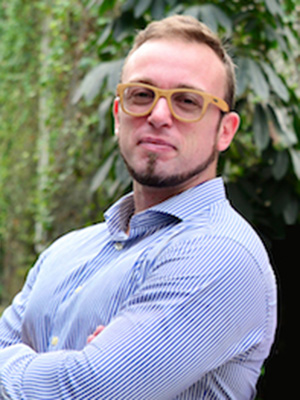}}]{Erik Cambria} Erik Cambria is an Associate Professor at Nanyang Technological University, Singapore, where he also holds the appointment of Provost Chair in Computer Science and Engineering. Erik Cambria is the founder of Erik Cambria is the Founder of SenticNet, a Singapore-based company offering B2B sentiment analysis services. Prior to joining NTU, he worked at Microsoft Research Asia (Beijing) and HP Labs India (Bangalore) and earned his PhD through a joint program between the University of Stirling and MIT Media Lab. His research focuses on the ensemble application of symbolic and subsymbolic AI to natural language processing tasks such as sentiment analysis, dialogue systems, and financial forecasting. Erik is recipient of many awards, e.g., the 2019 IEEE Outstanding Early Career Award, he was listed among the 2018 AI's 10 to Watch, and was featured in Forbes as one of the 5 People Building Our AI Future. He is Associate Editor of several journals, e.g., INFFUS, IEEE CIM, and KBS, Special Content Editor of FGCS, Department Editor of IEEE Intelligent Systems, and is involved in many international conferences as program chair and invited speaker. 
\end{IEEEbiography}




\end{document}